# Competition driven cancer immunoediting

Irina Kareva[1]

[1] Newman Lakka Institute, Floating Hospital for Children at Tufts Medical Center,

Boston, MA 02111

Irina.Kareva@tufts.edu

## Abstract

It is a well-established fact that tumors up-regulate glucose consumption to meet increasing demands for rapidly available energy by switching to purely glycolytic mode of glucose metabolism. What is often neglected is that cytotoxic cells of the immune system also have increased energy demands and also switch to pure glycolysis when they are in an activated state. Moreover, while cancer cells can revert back to aerobic metabolism, rapidly proliferating cytotoxic lymphocytes are incapable of performing their function when adequate resources are lacking. Consequently, in the tumor microenvironment there must exist competition for the common resources between cancer cells and the cells of the immune system, which may drive a lot of the tumor-immune dynamics. Proposed here is a model of tumor-immune-glucose interactions, formulated as a predator-prey-common resource type system, which allows to investigate possible dynamical behaviors that may arise as a result of competition for glucose, including tumor elimination, tumor dormancy and unrestrained tumor growth.

**Introduction**

Immunoediting is the process whereby the immune system can alter, or 'edit', the progression of tumor growth. It has been proposed that the process of immunoediting consists of three distinct phases: elimination phase, when the cytotoxic cells of the immune system reduce tumor size, equilibrium, when tumor size is maintained at a constant level, and escape, when the tumor circumvents the immune response, as tumor cells develop resistance and start growing again (1-3). The process of tumor escape is believed to be driven by mutations, and specifically, by the eventual appearance and subsequent selection for non-immunogenic tumor clones (2). We propose that circumvention of the immune system and subsequent tumor escape is not necessarily a result of increased genomic instability but can be explained solely by competition for common resources between cancer and immune cells in the tumor microenvironment.

It is a well-established fact that tumor cells can significantly up-regulate glucose consumption to meet the high energetic demands for cell survival and proliferation. Intriguingly, they frequently switch to fast but less efficient purely glycolytic mode of glucose metabolism even in the areas of ample oxygen supply, which yields 2 molecules of ATP compared to approximately 30 ATP that would have been obtained as a result of oxidative phosphorylation (4-6). Advantages of up-regulated glycolysis include not only significantly increased speed of access to the generated ATP but also the ability to divert glucose intermediates for new cell manufacturing (7; 8). A switch to pure glycolysis as a primary mode of glucose metabolism is often accompanied by up to 30-fold increase in glucose uptake and over-expression of glucose transporters, such as GLUT-1 and SGLT-1 (9) . Similarly, actively proliferating cytotoxic lymphocytes also switch to purely glycolytic mode of glucose metabolism, which is accompanied by a similar up-regulation of the expression of glucose transporters, such as GLUT-1 (10). Moreover, failure to increase glucose uptake during lymphocyte activation prevents cell growth and limits cell functionality, as cells deprived of adequate glucose supply cannot produce cytokines such as IFN-γ, which are required for effector function (11; 12). Therefore, it is not unreasonable to assume that in the tumor microenvironment there can exist competition for glucose between tumor and immune cells, and the outcome of this interaction would be determined by whichever cell type is the first to acquire access to glucose. This hypothesis relies specifically on the fact that T cells proliferate not only in the periphery but also in the tumor microenvironment (Thompson et al. 2010).

These considerations allow to conceive of the following theoretical set-up: as the primary tumor grows, the cells inside it switch to glycolysis due to oxygen deprivation, establishing a glycolytic core, while the cells on the outside of the tumor may still continue using more energy efficient aerobic metabolism. Assuming sufficient tumor immunogenicity and functional immune response, actively proliferating immune cells that are attracted to the tumor cite are expected to be able to succeed in competition for glucose at this stage of the tumor-immune interaction, since, unlike aerobic cancer cells that they are coming in contact with, immune cells are using purely glycolytic mode of metabolism, which, as was pointed out above, is accompanied by significant upregulation of nutrient transporters. As a result of successful tumor 'contraction' by the immune cells, glycolytic core of the tumor could become exposed, and these glycolytic tumor cells can now actively compete with the immune cells for the available glucose. Should the immune cells not be able to succeed (if, for instance, selection for extremely up-regulated nutrient transporters was more severe in the oxygen-deprived tumor core than it was for the immune cells in their environment), then they will not be able to undergo clonal expansion, thus allowing the tumor to circumvent the immune system and continue growing. A schematic representation of the proposed scenario is given in Figure 1.

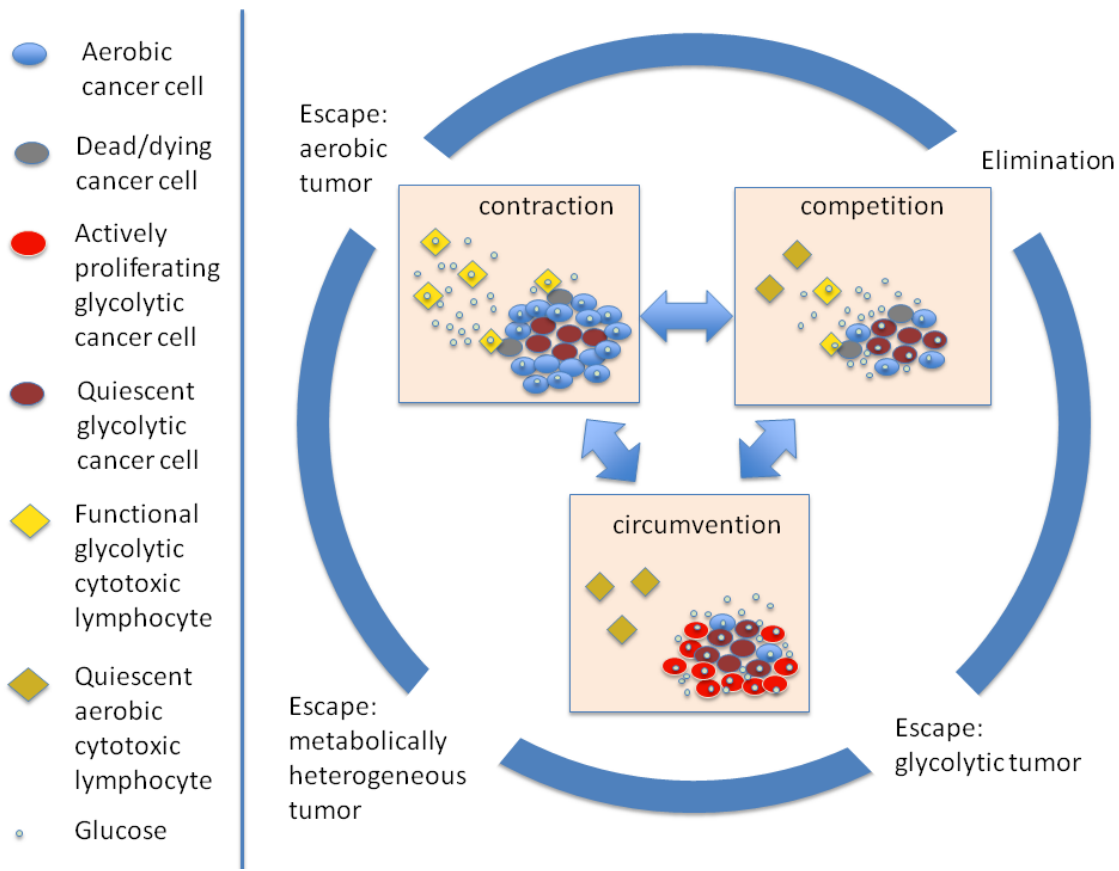

Figure 1. Proposed scenario of metabolism-driven tumor escape. In the tumor microenvironment, glucose is taken up by cytotoxic lymphocytes that use glycolysis as a primary mode of metabolism, causing the tumor to contract and exposing its glycolytic core. The glycolytic tumor cells have up-regulated nutrients transporters, thus posing competition to immune cells for resources. Lymphocytes cannot proliferate unless their nutrient demands are met, which can allow tumor cells to circumvent immune response, leading to tumor escape.

We introduce a mathematical model in order to investigate whether such a behavior can indeed be realized using the most basic set of assumptions about tumor-immune-glucose interactions, as well as to study any other possible regimes that may arise in such a system. The proposed model is a 4-dimensional system of ordinary differential equations of the form predator-prey-common resource, where the cytotoxic lymphocytes are the predator, cancer cells are the prey, and glucose is the common resource. We hypothesize that depending on glucose availability we'll be able to observe the escape phase of immunoediting, which will be driven purely by competition for resources in the tumor microenvironment.

**Model Description**

Proposed here is a conceptual mathematical model aimed to describe the dynamical interactions between a heterogeneous population of tumor cells (prey) and cytotoxic immune cells (predator), competing for glucose in the tumor microenvironment (the common resource).

The population of tumor cells $T(t)$ is divided into two subpopulations of aerobic and glycolytic cells such that $T(t) = T_a(t) + T_g(t)$. The dynamics of the two subpopulations is described by identical equations that differ solely in intrinsic parameter values. Tumor growth is assumed to be proportional to the amount of glucose that is available to the cells in the microenvironment, which is described by the term $r_j T_j(t) \dfrac{G(t)}{1 + b_T G(t)}$, where *j=a* represents aerobic cells and *j=g* represents glycolytic cells; this functional form was taken from classical models of consumer-resource interactions, reviewed in Cantrell et al. (13). Tumor cells can die either naturally proportionally to $\mu_j T_j(t)$ or they can be killed by cytotoxic lymphocytes $I(t)$, which is accounted for by the term $e_j T_j(t) \frac{I(t)/T_j(t)}{s + I(t)/T_j(t)}$; this functional form is taken from models describing tumor-immune interactions developed by de Pillis et al. (14; 15).

Glucose $G(t)$ is replenished from the blood stream at a constant rate $G_0$ and is consumed differentially by tumor cells according to functional form $d_j T_j(t) \frac{G(t)}{1 + b_T G(t)}$ and by immune cells $d_I I(t) \frac{G(t)}{1 + b_I G(t)}$; since actively proliferating cytotoxic immune cells switch to glycolysis, we assume that $d_I \approx d_g$; the rate of glucose consumption by aerobic cells $r_a$ however has been estimated to be up to 20-30 times lower (9). Glucose is also consumed by normal tissue, which in our model is accounted for by the term $\mu_G \frac{G(t)}{1 + b_N G(t)}$. We assume here that $b_N \ll 1$, which allows to approximate this term as $\mu_G G(t)$.

Finally, cytotoxic immune cells are recruited to the tumor site at a rate $i_0 T(t)$; parameter $i_0$ can vary significantly depending on the degree of systemic immune stimulation or tumor-induced suppression; it can also be affected by various types of immunotherapy. Immune cells are assumed to die at some natural rate $\mu_I I(t)$, which can also vary depending both on overall health of the patient and on various external stimuli, such as low-dose irradiation (16). Finally, expansion of cytotoxic immune cells depends both proportionally to the number of tumor cells that have already been killed by the immune cells (14; 15), and proportionally to the amount of glucose available to the immune cells in tumor microenvironment. This restriction is imposed to account for the fact that due to their high energy demands, even fully activated cytotoxic immune cells cannot perform their function if they do not have access to sufficient amounts of glucose (10).

The assumptions that were used in model formulation can be summarized as follows:

1. Tumor population consists of aerobic and glycolytic cells.
2. Glycolytic tumor cells grow faster but also deplete common resources faster (they have higher nutrient uptake rates).
3. Glucose is depleted at different rates by tumor and immune cells; aerobic cells deplete it slower, glycolytic tumor cells and immune cells deplete it faster.
4. Population of immune cells decays in the absence of tumor, since cytotoxic lymphocytes need co-stimulatory signals from tumor cells to enable them to take up glucose; otherwise they die by starvation.
5. Expansion and functionality of immune cells is glucose-dependent: they cannot grow and perform their function if they do not have access to sufficient amounts of glucose.

Taking all of these assumptions into account, we end up with the following system of ordinary differential equations:

$$\frac{dT_a}{dt} = \underbrace{r_a T_a(t)\frac{G(t)}{1+b_T G(t)}}_{\text{glucose dependent growth}} - \underbrace{e_a \frac{T_a(t)I(t)}{I+s(T_a(t)+T_g(t))}}_{\text{ratio-dependent death by CTLs}} - \underbrace{m_a T_a(t)}_{\text{natural death}}$$

$$\frac{dT_g}{dt} = r_g T_g(t)\frac{G(t)}{1+b_T G(t)} - e_g \frac{T_g(t)I(t)}{I(t)+s(T_a(t)+T_g(t))} - m_g T_g(t)$$

$$\frac{dG}{dt} = \underbrace{(G_0 - m_G G(t))}_{\text{natural G inflow/outflow}} - \underbrace{(d_a T_a(t)+d_g T_g(t))\frac{G(t)}{1+b_T G(t)}}_{\text{G consumed by the tumor}} - \underbrace{d_I I(t)\frac{G(t)}{1+b_I G(t)}}_{\text{G consumed by immune cells}}$$

$$\frac{dI}{dt} = (\underbrace{i_0(T_a(t)+T_g(t))}_{\text{systemic immune stimulation}} - \underbrace{m_I I(t)}_{\text{natural decay}}) + \underbrace{r_I I(t)\left(\frac{G(t)}{1+b_I G(t)}\right)\left(\frac{I(t)(T_a(t)+T_g(t))}{I(t)+s(T_a(t)+T_g(t))}\right)}_{\substack{\text{cell expansion stimulated by debris from previously}\\\text{killed tumor cells and modulated by glucose}}} \quad (1)$$

Description of the parameters and sample values are given in Table 1.

**Results**

The goal of the numerical computations conducted for this model is to investigate whether we can observe the theoretically predicted cycle of metabolism-modulated immune escape given the simplest set of assumptions outlined in Figure 1, as well as to see what some of the other possible dynamical regimes there can be. We chose to focus on the effect of parameters $i_0$ (systemic immune stimulation), $e_j$ (effectiveness of tumor elimination by the immune system, $j = a, g$) and $m_I$ (death rate of cytotoxic lymphocytes) on system dynamics because we believe that these parameters can, at least in theory, be manipulated in an experimental setting via various therapeutic interventions, such as chemo-, radiation and immunotherapy.

*Aerobic and glycolytic tumor cells can cooperate to defeat the immune system*

In the first set of simulations, we varied parameter $i_0$ and were indeed able to observe the theoretically predicted cyclical dynamics. Other parameters were held constant at $b_I = b_T = 0.9$, $e_a = e_g = 0.1$, $r_a = 0.1$, $r_g = 1.9 r_a$, $s = 0.7$, $d_a = 0.1$, $d_g = d_I = 2 d_a$, $m_a = 0.01$, $m_g = 3 m_a$, $G_0 = 1$, $m_I = 0.01$, $r_I = 0.01$, $T(0) = I(0) = C(0) = 1$, $T_a(0) = 0.1 T$. Parameter values were chosen based on the values provided in the validated mathematical models proposed by de Pillis et al. (14; 15); in other cases, parameter values were estimated based on theoretical considerations. It is important to emphasize that the primary purpose of the proposed model is to validate theoretical predictions, identify qualitative regimes that are possible within the described framework and to propose further research directions rather than to match predictions to specific pre-existing data.

As we progressively increased parameter $i_0$, we were able to observe the dynamical shift from tumor persistence (Figure 2a) to multiple cyclical regimes (Figure 2b) to decrease in cycle number before tumor size stabilized at a low equilibrium value (Figure 2c) to equilibrium state achieved without any preceding cyclical dynamics (Figure 2d).

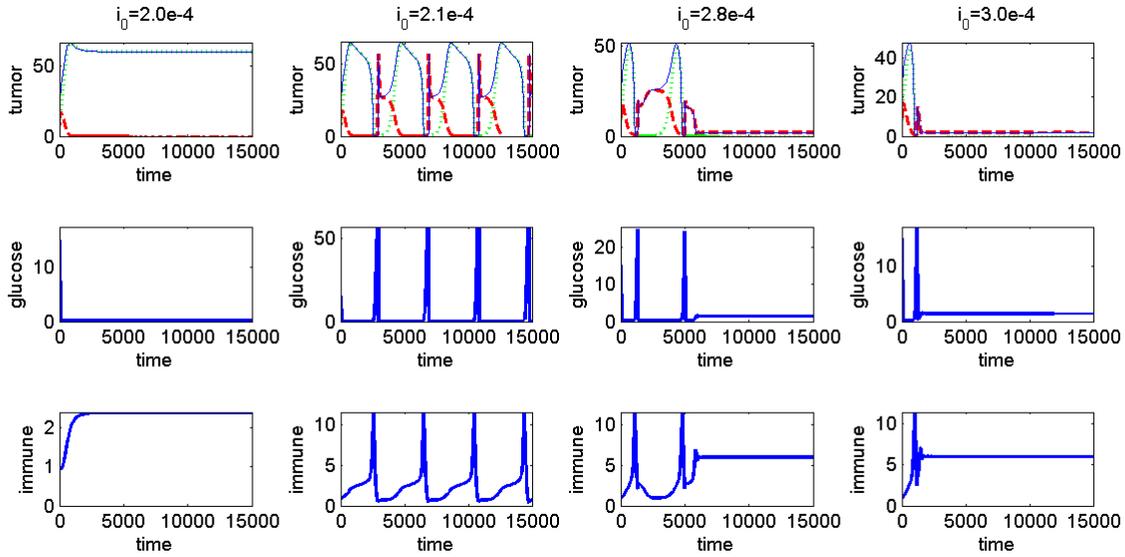

Figure 2. Different dynamical regimes that can be observed through progressively increasing parameter $i_0$, which represents systemic immune stimulation. Some possible regimes that can be observed range from a) escape of a primarily aerobic tumor to b) sustained cyclical behavior described in Figure 1 to c) several iterations of cyclical behavior before bringing down tumor size to d) tumor control.

Within this set of simulations we could observe that predicted cyclical dynamics can recur several times before the immune system brings down the overall tumor size. Moreover, as one can see, the composition of the tumor with respect to metabolic strategy can be different at different stages of the cycle: while in Figure 2a the tumor is composed primarily of aerobic cells, in Figure2c and d, the tumor is composed of glycolytic cells.

The specifics of the dynamics that drive the oscillations can be observed in more detail in Figure 3a, which is a magnification of the dynamics shown in Figure2b:

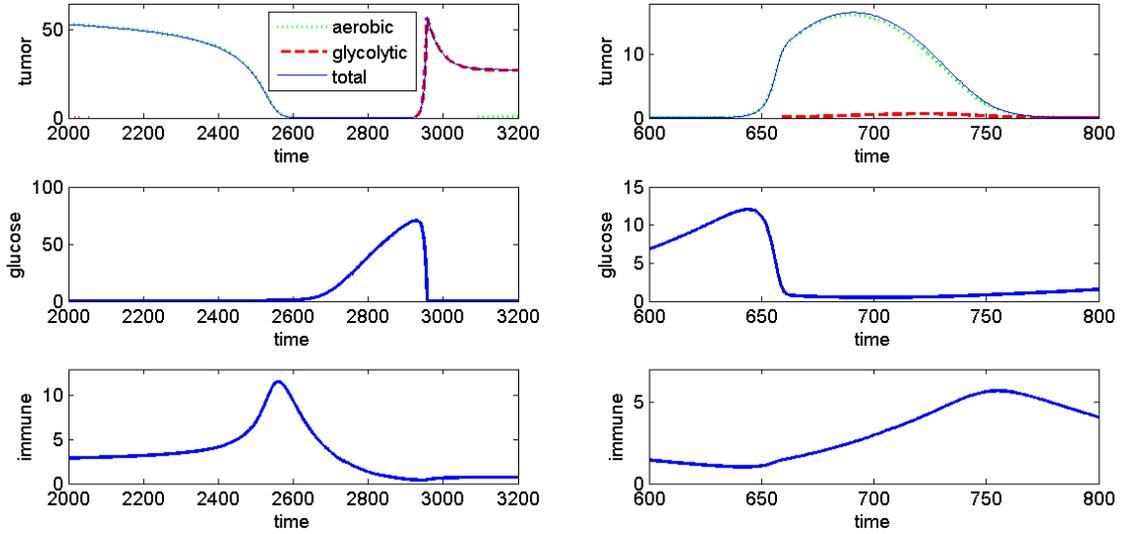

Figure 3. Details underlying the dynamics of a) oscillations driven by glycolytic cancer cells ($b_T = 0.9$, $e_j = 0.1$, $r_g = 1.8 r_a, i_0 = 2.1e-4$) and b) oscillations driven by aerobic cancer cells ($b_T = 0.2$, $e_j = 0.2$, $r_g = 2.8 r_a$, $i_0 = 2.0e-3$). The first case corresponds to competition driven circumvention of immune surveillance; the second case corresponds to a variant of a classic predator-prey type system.

We can observe that increase in the number of immune cells corresponds to decrease in the number of aerobic tumor cells. As tumor size decreases, so does the population of immune cells, since con-stimulatory molecules from tumor debris are necessary for clonal expansion of immune cells (this is accounted for by the last term in the equation for dL/dt). As the population of immune cells decreases and stops utilizing glucose, its levels rise and it can now be then taken up by expanding glycolytic cells. This causes a sharp drop in glucose levels in the tumor microenvironment, preventing immune cells from proliferating, in this case not from lack of co-stimulation but from lack of resources necessary for clonal expansion.

In such a scenario, the dynamics can be interpreted as cooperation between aerobic and glycolytic tumor cells against cytotoxic lymphocytes: glycolytic cells take up the resources, allowing the tumor to circumvent the immune system and for the more successful of the remaining tumor cells to grow (noticeably, these could be either aerobic or glycolytic cancer cells; however, the circumvention of the immune system is in this case modulated by glycolytic cancer cells).

The dynamics observed under these parameter values corresponds to the circular dynamics predicted in Figure 1.

*Increased immune stimulation can be sufficient to control tumor size but not to eliminate it*

Next, we evaluated whether further increases in overall immune stimulation could lead to permanent tumor elimination; a positive answer would suggest that tumor growth could in some cases be controlled through immunotherapy alone. We observed that increasing the value of $i_0$ by five orders of magnitude did not lead to tumor elimination; it allowed destabilizing the tumor, yielding oscillatory regimes of a much smaller magnitude than in the previous set of simulations, but complete elimination was not observed (see Figure 4).

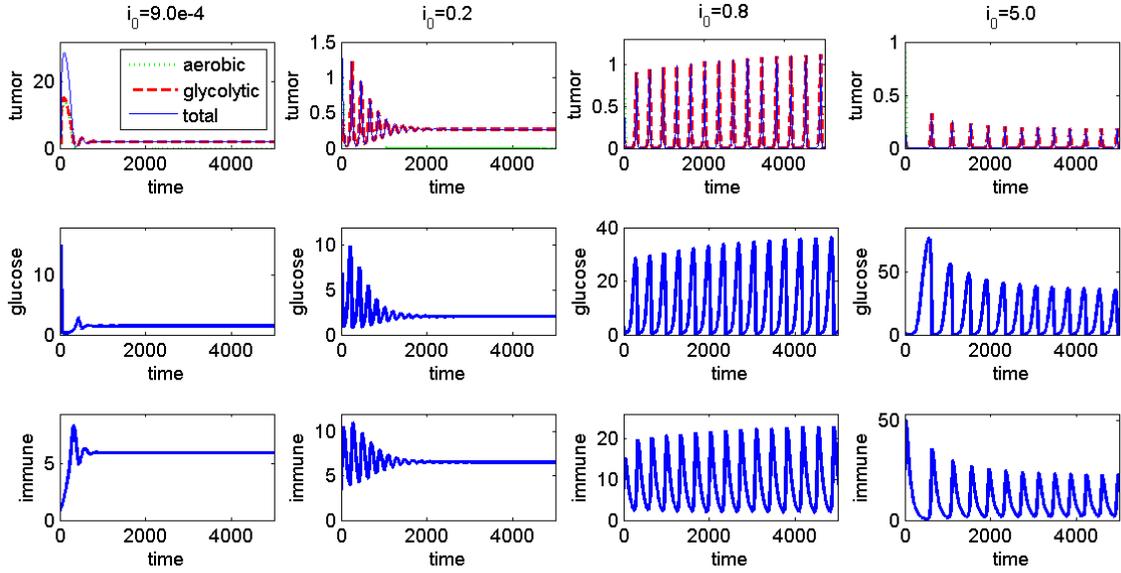

Figure 4. Effects of further increases of parameter $i_0$. The simulations suggest that increased immune stimulation can decrease the amplitude of oscillations but cannot yield complete tumor elimination.

We then hypothesized that even with strong supplementary immune stimulation, tumor elimination can be achieved only if the parameter of tumor elimination $e_j$ is large enough (out of parameters that we are choosing to focus on in this paper as the ones that can be manipulated in a clinical setting). For this set of simulations, we fixed $i_0 = 2.1e-4$ and then progressively increased parameter $e_j$ (see Figure 5).

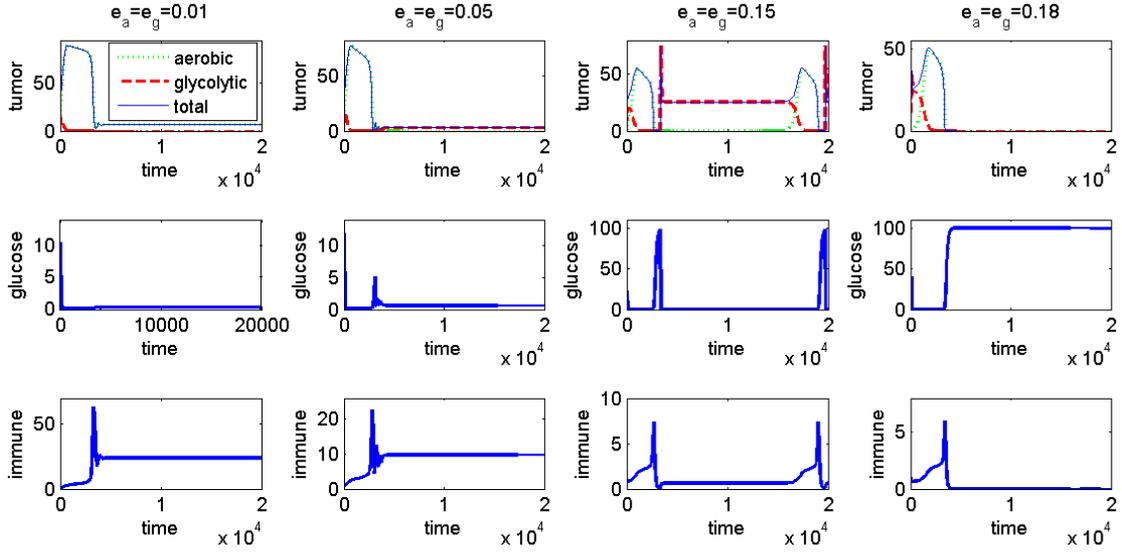

Figure 5. Effects of increases of parameter of tumor elimination $e_j$, $j = a, g$ while keeping parameter $i_0 = 2.1e - 4$. Calculations suggest that sufficient increases in the value of $e_j$ can indeed lead to full tumor elimination.

These sets of simulations demonstrate that predictably, high enough rate of killing of tumor cells is necessary for tumor elimination. However, if parameter $e_j$ cannot be manipulated, or increased to a large enough value, higher degree of tumor control would be achieved through sufficient immune stimulation, even though it is not expected to completely eliminate the tumor regardless of the level of stimulation.

*High death rates of immune cells can promote tumor growth*

Next, we evaluated the effects of high death rates of immune cells on overall system dynamics, since cytotoxic therapeutic interventions have adverse effects not only on the tumor cells but also on healthy tissues, including cells of the immune system (Mackall et al., 1997). We conducted numerical experiments from the area of phase-parameter space where tumor was completely eliminated and then observed the effects of increases of parameter $m_I$ on overall system dynamics. Parameters were taken to be $b_T = 0.9$, $i_0 = 0.0001$, $r_g = 2.8r_a$, $e_a = e_g = 0.1$; all other parameter values remained the same.

We observed that progressively increasing the values of parameter $m_I$ lead to eventual rapid tumor escape, preceded by progressively increasing oscillatory regimes of small amplitude (see Figure 6). These results suggest that not only should effects of cytotoxic therapies on the immune cells be taken into account but also that immune-modulated tumor control can be disrupted if the patient is exposed to anything that could severely increase mortality of cytotoxic lymphocytes.

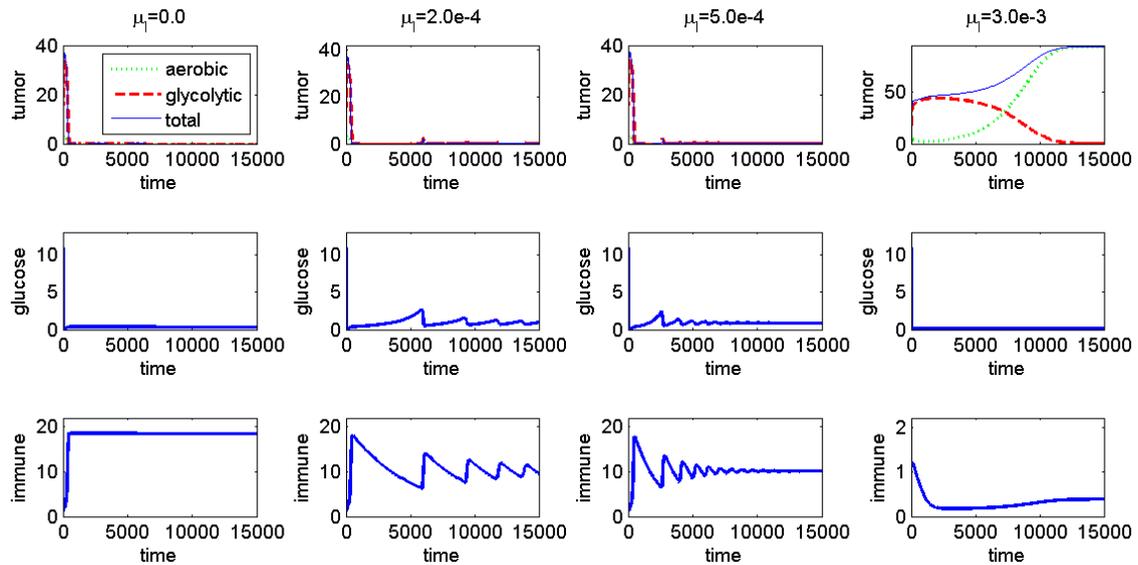

Figure 6. This set of calculations illustrates the effect that increased mortality of immune cells can have on over all tumor dynamics, ranging from complete elimination (set a) to increased destabilizing oscillatory behavior (sets b and c) to tumor escape (set d). Other parameters are held constant at $i_0 = 1.0e-4$, $b_T = 0.9$, $r_g = 2.8r_a$, $e_a = e_g = 0.1$.

**Discussion**

It is conventionally believed that the process of tumor escape from the immune system is driven by the eventual mutation induced appearance of non-immunogenic cell clones, which allow the cancer to progress unrecognized by the immune system. We propose that genomic instability of tumor cells is not imperative for immune escape, which can be mediated solely by competition for resources, and specifically glucose, between tumor and immune cells in the tumor microenvironment.

Actively proliferating cells, whether it be tumor or immune cells, up-regulate glycolysis as a primary mode of glucose metabolism, both due to its speed compared to more efficient but slower oxidative phosphorylation, and due to its ability to provide rapidly dividing cells with glycolytic intermediates necessary for biosynthesis of nucleic acids (8; 10). This also requires glycolytic cells to overexpress nutrient transporters, such as GLUT-1, 20-40 fold compared to aerobic cells, in order to enable the cells to meet their energy demands (9; 10). Therefore, it is not unlikely that in the tumor microenvironment there must exist some level of competition for glucose between actively proliferating cancer and immune cells.

Such theoretical considerations allow to predict the following dynamics: as the tumor grows, cells it its core become oxygen deprived and are forced to switch to glycolysis, while the cells on the outer rim of the tumor continue using oxidative phosphorylation as a more efficient way to obtain energy. If the aerobic cells on the tumor rim are sufficiently immunogenic to initiate a strong immune response, then cytotoxic lymphocytes eliminate these cells first, eventually exposing the glycolytic core.

Unlike the aerobic tumor cells, glycolytic tumor cells can now present sufficiently strong competition to immune cells for glucose, since even the most potent immune cells lose the ability to perform their function in the state of nutrient deprivation (11). As a result, glycolytic tumor cells can hinder the activity of the immune system solely via competition for glucose, allowing tumor escape. This mechanism is also summarized in Figure 1.

(Interestingly, there exists some evidence of another possible mechanism of cooperation between aerobic and glycolytic cells in the tumor, where glycolytic cancer cells provide aerobic cells with lactate which they can then utilize as part of the citric acid cycle (17); however, this consideration is not taken into account in the current version of the model).

In order to investigate the variety of possible outcomes of glucose-driven competitive interactions between tumor and immune cells, we formulated a mathematical model of the type 'predator-prey-common resource', where the predator (immune cells) competes with the prey (aerobic and glycolytic tumor cells) for the common resources (glucose).

The proposed 4-dimensional system of ordinary differential equations was solved numerically using Matlab2010b. For calculating numerical solutions we focused on varying parameters that, at least in theory, can be manipulated in an experimental setting, and specifically parameter of overall immune stimulation $i_0$, parameter of the effectiveness of tumor elimination by immune cells $e$ and parameter of immune cell mortality $m_I$; all the parameter values used are given in the text.

First we varied parameter $i_0$, which in our model represents systemic immune stimulation and which in an experimental setting could be influenced positively by immunotherapy or negatively by tumor-induced immune suppression. We indeed were able to observe the theoretically predicted cyclical dynamics (see Figure 2). As tumor size decreased, so did the population of immune cells, since co-stimulatory molecules from tumor debris were required for clonal expansion of immune cells. As the population of immune cells decreased, glucose levels rose, which made it available to the expanding population of glycolytic tumor cells. This in turn caused a sharp drop in glucose levels in the tumor microenvironment, preventing immune cells from proliferating. The details of this interaction are visualized on Figure 3a. Moreover, we showed that such cyclical dynamics could in fact occur several times before the tumor either shrinks or escapes.

Next, we wanted to investigate whether the tumor can be eliminated through increased stimulation of the immune system alone, i.e., whether one could expect to cure the disease solely via extensive immune stimulation, such as through immunotherapy. We observed that increasing immune stimulation by five orders of magnitude still did not yield complete tumor elimination. Only sufficiently increased killing of the tumor cells by the immune system was shown to lead to complete tumor elimination (see Figure 5). These results suggest that immunotherapy could be an effective complementary therapy but it should not normally be expected to cure the cancer.

Finally, we evaluated the impact of high death rates of immune cells on overall tumor dynamics, since this parameter can be influenced both acutely as a side effect of cytotoxic therapies (18-21) and chronically via environmental factors, such as low-dose radiation (unpublished data obtained in our lab); some studies also suggest that lymphocyte count recovery after anti-cancer treatment with cytotoxic drugs can serve as a predictor of superior survival in AML and ALL patients (22-24). As expected, we observed that progressively increasing parameter of immune cell mortality lead to eventual tumor escape, preceded by progressively increasing oscillatory regimes of small amplitude (see Figure 6), suggesting that immune cell mortality could be another dimension along which predicted therapy efficacy should be evaluated.

From the point of view of therapeutic implications, the proposed theoretical model of glucose-modulated immunoediting can provide an explanation for limited success of immunotherapy as a mode of treatment (25; 26).: even activated cytotoxic lymphocytes, such as those obtained via adoptive T cell transfer, would not be able to undergo clonal expansion and perform their function if glycolytic tumor cells consume the resources from the tumor microenvironment. At the same time, vaccine therapies, which have shown most promise, provide systemic immune stimulation, which, as the model predicts, can be of crucial importance for maintaining tumor size at a controlled level (27; 28).

More generally, the proposed theoretical construct provides an additional explanation for persistence of fast but inefficient aerobic glycolysis observed in many tumors, a phenomenon also known as Warburg effect (5). Not only can up-regulated glycolysis provide building materials for actively growing cells (8) and promote successful competition with somatic cells through creating an acidic microenvironment (29-31) but it can also allow tumor cells to circumvent the immune system via competition for nutrients.

Intriguingly, from this point of view, very strong immune response could in fact promote tumor progression, since these cells would be more efficient at exposing the tumor's glycolytic core. Therefore, it is the moderately strong immune stimulation that would be able to most successfully control tumor growth. This prediction would correspond to predictions made by both in silico and in vivo experiments conducted by Gatenby et al. (Cancer Res 2009), where the authors demonstrate that a therapeutic set-up, in which chemotherapeutic treatment was continuously modulated to achieve a fixed tumor population, allowed mice to coexist with the tumors indefinitely. The authors suggested that best results in the long term could be achieved by permitting a significant population of chemo-sensitive cells to survive so that they, in turn, suppress proliferation of the less fit but chemo-resistant subpopulations (32). Moderately but not excessively effective immune response would present a natural example of such an adaptive therapy, which can help better understand and improve existing immunotherapies.

**Acknowledgements**

This paper is an abridged version of the following manuscript: Kareva I., Beheshti, A., L. Hlatky and P. Hahnfeldt. "Cancer Immunoediting: A Process Driven by Metabolic Competition." The author would like to sincerely thank her mentor Philip Hahnfeldt for his mentorship during this project. The author would also like to thank Heiko Enderling and Kathleen Wilkie for extremely helpful suggestions and discussions.

This research was supported by the Office of Science (BER), U.S. Department of Energy, under Award Number DE-SC0001434 (to Philip Hahnfeldt, IK's post-doctoral mentor) and Award Number DE-SC0002606 (to Lynn Hlatky), and by a National Cancer Institute grant ICBP 1U54CA149233 (to Lynn Hlatky). The research has been conducted during IK's postdoctoral training at Center of Cancer Systems Biology, St. Elizabeth's Medical Center, Boston, MA, 02135, in 2012-2013.

**Bibliography**

1. Dunn GP, Bruce AT, Ikeda H, Old LJ, Schreiber RD. Cancer immunoediting: from immunosurveillance to tumor escape. Nat Immunol. 2002 Nov;3(11):991-8.

2. Dunn GP, Old LJ, Schreiber RD. The three Es of cancer immunoediting. Annu Rev Immunol. 2004;22:329-60.

3. Kim R, Emi M., Tanabe K. Cancer immunoediting from immune surveillance to immune escape. Immunology. 2007 May;121(1):1-14.

4. Kim JW, Dang CV. Cancer's molecular sweet tooth and the Warburg effect. Cancer Res. 2006 Sep 15;66(18):8927-30.

5. Warburg O. On the origin of cancer cells. Science. 1956 Feb 24;123(3191):309-14.

6. Semenza GL. A return to cancer metabolism. J Mol Med (Berl). 2011 Mar;89(3):203-4.

7. Vander Heiden MG, Cantley LC, Thomspon CB. Understanding the Warburg effect: the metabolic requirements of cell proliferation. Science. 2009 May 22;324(5930):1029-33.

8. Weinberg F, Chandel NS. Mitochondrial metabolism and cancer. Ann N Y Acad Sci. 2009 Oct;1177:66-73.

9. Ganapathy V, Thangaraju M, Prasad PD. Nutrient transporters in cancer: relevance to Warburg hypothesis and beyond. Pharmacol Ther. 2009 Jan;121(1):29-40.

10. Fox CJ, Hammerman PS, Thompson CB. Fuel feeds function: energy metabolism and the T-cell response. Nat Rev Immunol. 2005 Nov;5(11):844-52.

11. MacIver NJ, Jacobs SR, Wieman HL, Wofford JA, Coloff JL, Rathmell JC. Glucose metabolism in lymphocytes is a regulated process with significant effects on immune cell function and survival. J Leukoc Biol. 2008 Oct;84(4):949-57.

12. Jacobs SR, Herman CE, MacIver NJ, Wofford JA, Wieman HL, Hammen JJ, et al. Glucose uptake is limiting in T cell activation and requires CD28-mediated Akt-dependent and independent pathways. J Immunol. 2008 Apr 1;180(7):4476-86.

13. Cantrell RS, Cosner C, Shigui R. Intraspecific interference and consumer-resource dynamics. Discrete Contin Dyn Syst Ser B. 2004;4:527–546.

14. de Pillis LG, Radunskaya AE, Wiseman CL. A validated mathematical model of cell-mediated immune response to tumor growth. Cancer Res. 2005 Sep 1;65(17):7950-8.

15. de Pillis LG, Gu W, Radunskaya AE. Mixed immunotherapy and chemotherapy of tumors: modeling, applications and biological interpretations. J Theor Biol. 2006 Feb 21;238(4):841-62.

16. Jahns J, Anderegg U, Saalbach A, Rosin B, Patties I, Glasow A, et al. Influence of low dose irradiation on differentiation, maturation and T-cell activation of human dendritic cells. Mutat Res. 2011 May 10;709-710:32-9.

17. Feron O. Pyruvate into lactate and back: from the Warburg effect to symbiotic energy fuel exchange in cancer cells. Radiother Oncol. 2009 Sep;92(3):329-33.

18. Stahnke K, Fulda S, Friesen C, Strauss G, Debatin KM. Activation of apoptosis pathways in peripheral blood lymphocytes by in vivo chemotherapy. Blood. 2001;98(10):3066–3073.

19. Mackall CL, Fleisher TA, Brown MR, et al. Lymphocyte depletion during treatment with intensive chemotherapy for cancer. Blood. 2001 Nov 15;98(10):3066-73.

20. Mackall CL, Fleisher TA, Brown MR, Andrich MP, Chen CC, Feuerstein IM, et al. Distinctions between CD8+ and CD4+ T-cell regenerative pathways result in prolonged T-cell subset imbalance after intensive chemotherapy. Blood. 1997 May 15;89(10):3700-7.

21. Rübe CE, Grudzenski S, Kühne M, Dong X, Pief N, Löbrich M, et al. DNA double-strand break repair of blood lymphocytes and normal tissues analysed in a preclinical mouse model: implications for radiosensitivity testing. Clin Cancer Res. 2008 Oct 15;14(20):6546-55.

22. De Angulo G, Yuen C, Palla S, Anderson PW, Zweidler-McKay PA. Absolute lymphocyte count is a novel prognostic indicator in ALL and AML: implications for risk stratification and future studies. Cancer. 2008 Jan 15;112(2):407-15.

23. Siddiqui M, Ristow K, Markovic SN, Witzig TE, Habermann TM, Colgan JP, et al. Absolute lymphocyte count predicts overall survival in follicular lymphomas. Br J Haematol. 2006 Sep;134(6):596-601.

24. Behl D, Porrata LF, Markovic SN, Letendre L, Pruthi RK, Hook CC, et al. Absolute lymphocyte count recovery after induction chemotherapy predicts superior survival in acute myelogenous leukemia. Leukemia. 2006 Jan;20(1):29-34.

25. Lesterhuis WJ, Haanen JB, Punt CJ. Cancer immunotherapy--revisited. Nat Rev Drug Discov. 2011 Aug 1;10(8):591-600.

26. Hoos A, Eggermont AM, Janetzki S, Hodi FS, Ibrahim R, Anderson A, et al. Improved endpoints for cancer immunotherapy trials. J Natl Cancer Inst. 2010 Sep 22;102(18):1388-97.

27. Pardoll D, Drake C. Immunotherapy earns its spot in the ranks of cancer therapy. J Exp Med. 2012 Feb 13;209(2):201-9.

28. Mellman I, Coukos G, Dranoff G. Cancer immunotherapy comes of age. Nature. 2011 Dec 21;480(7378):480-9.

29. Gatenby RA, Gillies RJ. Why do cancers have high aerobic glycolysis? Nat Rev Cancer. 2004 Nov;4(11):891-9.

30. Gillies RJ, Robey I, Gatenby RA. Causes and consequences of increased glucose metabolism of cancers. J Nucl Med. 2008 Jun;49 Suppl 2:24S-42S.

31. Robey IF, Baggett BK, Kirkpatrick ND, Roe DJ, Dosescu J, Sloane BF, et al. Bicarbonate increases tumor pH and inhibits spontaneous metastases. Cancer Res. 2009 Mar 15;69(6):2260-8.

32. Gatenby RA, Silva AS, Gillies RJ, Frieden BR. Adaptive therapy. Cancer Res. 2009 Jun 1;69(11):4894-903.

Table 1. Variables and parameters used in System 1. Variables are cast in units of biomass; rates are cast in units 1/time.

|  | Description | Sample value | Source |
|---|---|---|---|
| $T(t)$ | Population of tumor cells, composed of aerobic $T_a(t)$ and glycolytic $T_g(t)$ cells | $T(t) \geq 0$ | est. |
| $G(t)$ | Glucose | $G(t) \geq 0$ | est. |
| $I(t)$ | Cytotoxic immune cells | $I(t) \geq 0$ | est. |
| $r_a$ | Growth rate of aerobic cancer cells | $4.31 \times 10^{-1}$ | De pillis06 |
| $r_g$ | Growth rate of glycolytic cancer cells | $r_a \times k_1, k_1 \in [0.5, 5.0]$ | est. |
| $\mu_a$ | Natural death rate or aerobic cancer cells | 0.01 | est. |
| $\mu_g$ | Natural death rate of glycolytic cancer cells | $\mu_g \times k_2, k_2 \in [1.0, 10.0]$ | est. |
| $e_a$ | Fractional aerobic cell kill by immune cells | 0.1 | est. |
| $e_g$ | Fractional glycolytic cell kill by immune cells | $e_a - k_3, k_3 \in [-0.1, 0.1]$ | est. |
| $s$ | Steepness coefficient of the tumor cell kill by immune cells | $6.18 \times 10^{-1}$ | dePillis05 |
| $G_0$ | Rate of glucose inflow into tumor microenvironment from blood stream | $G_0 \in [0.1, 10.0]$ | est. |
| $\mu_g$ | Rate of glucose outflow from tumor microenvironment from blood stream | 0.01 | est. |
| $d_a$ | Rate of glucose consumption by aerobic tumor cells | 0.1 | est. |
| $d_g$ | Rate of glucose consumption by glycolytic tumor cells | $d_a \times k_4, k_4 \in [1.0, 30.0]$ | Ganapathy09 |
| $d_I$ | Rate of glucose consumption by the immune cells | $d_I \approx d_a$ | Fox05 |
| $i_0$ | Rate of tumor-stimulated inflow of immune cells | $i_0 \in [.001, 0.100]$ | n/a |
| $\mu_I$ | Death rate of immune cells | $2.0 \times 10^{-2}$ | dePillis05 |
| $r_I$ | Rate at which immune cells are stimulated to be produced as a result of coming in contact with tumor debris from previously killed tumor cells | $1.13 \times 10^{-1}$ | De pillis05 |